# Radiological and Biological Dictionary of Radiomics Features: Addressing Understandable AI Issues in Personalized Breast Cancer; Dictionary Version BM1.0

Arman Gorji[1], Nima Sanati[1], Amir Hossein Pouria[2], Somayeh Sadat Mehrnia[3], Ilker Hacihaliloglu[4,5], Arman Rahmim[5,6], Mohammad R. Salmanpour[5,6,7*]

[1] NAIRG, Department of Neuroscience, Hamadan University of Medical Sciences, Hamadan, Iran
[2] Department of Computer Engineering, Amirkabir University of Technology, Tehran, Iran
[3] Department of Integrative Oncology, Breast Cancer Research Center, Motamed Cancer Institute, ACECR, Tehran, Iran
[4] Department of Medicine, University of British Columbia, Vancouver, BC, Canada
[5] Department of Radiology, University of British Columbia, Vancouver, BC, Canada
[6] Department of Integrative Oncology, BC Cancer Research Institute, Vancouver, BC, Canada
[7] Technological Virtual Collaboration (TECVICO Corp.), Vancouver, BC, Canada

(*) Corresponding Author: Mohammad R. Salmanpour, PhD; msalman@bccrc.ca

**ABSTRACT**

**Background:** Radiomics-based AI models show potential in breast cancer diagnosis but lack interpretability. This study bridges the gap between radiomic features (RF) and BI-RADS descriptors through a clinically interpretable framework.

**Methods:** We developed a dual-dictionary approach. First, a Clinically-Informed Feature Interpretation Dictionary (CIFID) was constructed by mapping 56 RFs to BI-RADS descriptors (shape, margin, internal enhancement) based on literature and expert review. Second, we applied this framework to a classification task to predict triple-negative (TNBC) versus non-TNBC subtypes using Dynamic Contrast-Enhanced MRI data from a multi-institutional cohort of 1,549 patients. We trained 27 machine learning classifiers with 27 feature selection methods. Using SHapley Additive exPlanations, we interpreted the model's predictions and developed a complementary Data-Driven Feature Interpretation Dictionary for 52 RFs, not included in the CIFID.

**Results:** The best-performing model (Variance Inflation Factor feature selector+Extra Trees Classifier) achieved an average cross-validation accuracy of 0.83±0.02. Our dual-dictionary approach successfully translated predictive RFs into understandable clinical concepts. For example, higher values of 'Sphericity', corresponding to a round/oval shape, were predictive of TNBC. Similarly, lower values of 'Busyness', indicating more homogeneous internal enhancement, were also associated with TNBC, aligning with existing clinical observations. The framework not only confirmed known imaging biomarkers but also identified novel, data-driven quantitative features.

**Conclusion:** This study introduces a novel dual-dictionary framework (BM1.0) that bridges RFs and the BI-RADS clinical lexicon. By enhancing interpretability and transparency of AI models, the framework supports greater clinical trust and paves the way for integrating RFs into breast cancer diagnosis and personalized care.

## 1. INTRODUCTION

Breast cancer is one of the leading causes of cancer-related mortality worldwide, accounting for 670,000 deaths in 2022 [1]. Early and accurate diagnosis is critical for improving patient outcomes, yet achieving precision in diagnosis and treatment remains a significant challenge. While artificial intelligence (AI) has shown promise in transforming breast cancer care, particularly in providing high precision in predictive models [2, 3, 4], its "black box" nature poses significant barriers to clinical adoption. The lack of transparency in AI models can undermine clinician trust and impede the translation of quantitative imaging insights into actionable clinical decisions, which are essential for personalized treatment plans [5, 6, 7, 8]. This issue is particularly pressing in the context of breast cancer, where early and accurate classification can significantly impact prognosis and survival rates [9].

In response to these challenges, Explainable AI (XAI) techniques, such as SHAP (Shapley Additive Explanations) and LIME (Local Interpretable Model-agnostic Explanations), have emerged as tools aimed at improving model interpretability. These methods provide post-hoc explanations that shed light on the importance of individual features in model predictions [10, 11]. Despite their potential, these techniques often fall short in translating abstract mathematical features into clinically meaningful terms that are intuitive for radiologists. Although XAI can highlight which features contribute most to a model's prediction, it does not bridge the gap between the model's abstract numerical outputs and the clinical language that radiologists use in daily practice [12, 13].

A central aspect of this challenge lies in the field of radiomics features (RF), which involves the extraction of high-dimensional, quantitative RFs from medical images [14, 15]. RFs have shown significant potential in revealing complex tumor characteristics such as spatial heterogeneity, texture patterns, and morphological complexity, which are often invisible to the naked eye [16, 17, 18, 19]. These features have been demonstrated to predict critical clinical factors, including molecular subtypes, treatment response, and overall prognosis [20, 21, 22, 23]. However, despite the power of these features, their abstract nature, which is often grounded in statistical and mathematical methods, creates a barrier to their clinical application, as they lack a direct, intuitive connection to the radiologist's established lexicon [24].

In clinical practice, the Breast Imaging Reporting and Data System (BI-RADS) provides radiologists with a standardized lexicon for reporting breast lesion characteristics, categorizing features such as shape, margin, and internal enhancement (IE) patterns into clear, actionable descriptors [25, 26]. These descriptors, such as "spiculated margin," "irregular shape," or "heterogeneous enhancement," form the foundation of clinical decision-making [27, 28], allowing for consistent communication and management recommendations, including biopsy and follow-up actions [29]. However, the inherently qualitative nature of BI-RADS assessments can lead to observer variability, and it remains limited in its ability to capture the full complexity of tumor characteristics.

Although both RFs and BI-RADS offer valuable insights into tumor characterization, they currently operate in disconnected domains. RFs provide quantitative, high-dimensional data, whereas BI-RADS relies on visual, qualitative descriptors. Several efforts [30, 31, 32, 33] have been made to link RFs to biological characteristics or to incorporate BI-RADS-like features into machine learning models, but the critical interpretability gap between these two domains remains largely unaddressed [34]. Existing XAI techniques, while useful for developers, typically provide explanations that remain abstract and disconnected from the familiar BI-RADS terminology, making it difficult for clinicians to integrate RF findings into their practice [35, 36].

The core research problem lies in the lack of a systematic mapping between the quantitative lexicon of RFs and the clinically intuitive BI-RADS descriptors. While RFs derived from imaging modalities like Dynamic Contrast-Enhanced MRI (DCE-MRI) have shown prognostic power in predicting key outcomes such as tumor subtype and treatment response [37], clinical professionals still struggle to interpret how specific RFs correspond to clinically meaningful characteristics. For instance, a model might predict malignancy based on a high Long Run High Gray Level from Gray Level Run Length Matrix Emphasis (GLRLM_LRHGLE) value, but clinical professionals may struggle to connect this mathematical feature to familiar BI-RADS descriptors like "spiculated margin" or "heterogeneous enhancement".

Thus, despite the significant potential of RFs, the disconnect between quantitative imaging features and clinical language hampers its broader adoption in routine clinical practice [38]. This study aims to bridge this gap by developing a Radiological and Biological Dictionary that links RFs extracted from breast MRI to the BI-RADS descriptors used by clinicians. By creating a structured, interpretable connection between RFs and BI-RADS, we aim to enhance the clinical applicability of RFs, making AI-driven models more transparent, interpretable, and actionable in the context of breast cancer diagnosis, especially considering past studies that found associations between specific BI-RADS descriptors and molecular subtypes [39].

To achieve this, we first create a structured table that connects each BI-RADS descriptor with its corresponding score, based on established guidelines. We then use expert input from physicians and RF data to build a comprehensive dictionary that enhances interpretability. To demonstrate the practical utility of this dictionary, we apply it to a classification task: predicting whether breast cancer is triple-negative (TNBC) or non-triple-negative (non-TNBC) using DCE-MRI data from 1,549 patients. Additionally, for RFs that were not initially included in the dictionary, we analyze their performance in the classification task to derive potential clinical meanings. By combining clinical expertise with data-driven insights to create a more interpretable connection between RFs and clinical descriptors, this approach aims to significantly enhance the clinical utility of RFs. Ultimately, our Radiological and Biological Dictionary will facilitate the integration of AI-driven RFs into routine breast cancer care, fostering collaboration between clinicians and data scientists, and leading to more personalized and precise diagnostic and treatment strategies.

## 2. MATERIALS AND METHODS

### 2.1. Exploring the Relationship Between BI-RADS Scoring System and Descriptors

Although the BI-RADS provides a standardized framework for classifying breast lesions, there is currently no definitive or universally quantified relationship between BI-RADS categories and individual imaging descriptors such as shape, margin, IE, Non-Mass Enhancement (NME), and kinetic curves. To better understand and characterize this relationship, we employed a structured literature search strategy to identify relevant peer-reviewed studies that explored how these descriptors correlate with BI-RADS categories and clinical outcomes. Based on the evidence gathered, we extracted key descriptors and their reported associations with malignancy risk and BI-RADS levels, which we systematically organized into a comparative table. This approach aims to provide a more reproducible, interpretable, and data-driven mapping between radiological semantics and BI-RADS categories, laying the groundwork for aligning RFs with clinically meaningful imaging traits and BI-RADS categories.

## 2.2. Construction of the Clinically-Informed Feature Interpretation Dictionary (CIFID)

After creating a mapping between radiological semantics and BI-RADS categories, we investigated the relationship of descriptors of BI-RADS score associated with different RFs. We used PyRadiomics [40], standardized in reference to the IBSI [41], with 108 standardized RFs as our reference RF set. These features encompassed 19 First-Order (FO), 15 Shape-based features (SF), 23 from the Gray Level Co-occurrence Matrix (GLCM), 16 from the Gray Level Size Zone Matrix (GLSZM), 16 from GLRLM, 5 from the Neighborhood Gray Tone Difference Matrix (NGTDM), and 14 from the Gray Level Dependence Matrix (GLDM). Supplemental File, Sheet 1, provides a comprehensive list of RFs, including their abbreviations and semantic definitions. By focusing on the semantic meaning of RFs and visual assessment features derived from BI-RADS scoring system, we have developed a conceptual method to map the descriptive language as used by radiologists and clinicians to the quantitative data extracted from imaging studies. This approach not only enhances the explainability and interpretability of AI models by addressing the relationship between RFs and model outputs, but also improves the clinical relevance and usability of RFs in practice [32, 42, 43]. Furthermore, this dictionary was validated by three medical physicists and five medical professionals, including three experienced doctors of medicine, a radiologist, and a biologist who are familiar with RFs and AI analysis.

## 2.3. Understandable Classification Task

To demonstrate the practical utility of our dictionary, we designed a straightforward classification task to predict breast cancer molecular subtypes using RFs extracted from breast DCE-MRI. This task not only illustrated how the CIFID can be applied to interpret model outputs in a clinically meaningful manner but also helped identify and address gaps within the dictionary. For RFs not initially included in the CIFID, we conducted SHAP analysis on the classification model to assess the importance and directional influence of each feature. Based on the SHAP results, we developed a complementary Data-Driven Feature Interpretation Dictionary (DDFID), offering statistically informed interpretations for the remaining features. This dual approach allowed us to establish a more comprehensive and interpretable framework for linking quantitative RFs to clinically relevant descriptors.

***Data and Pre-processing**, and Molecular Subtype Prediction.* As shown in Table 1, we curated a multi-institutional dataset comprising six breast cancer cohorts: ISPY1 (166 patients), ISPY2 (980 patients), NACT (46 patients), and DUKE (287 patients) from the MAMA-MIA [44] consortium, as well as two external datasets, Advanced-MRI-Breast-Lesions (AMBL) (30 patients) [45] and The Cancer Genome Atlas Breast Invasive Carcinoma Collection (TCGA-BRCA) (40 patients) [46]. From each subject, we extracted the late post-contrast DCE-MRI sequence along with their corresponding tumor region of interest (ROI). As shown in Figure 1 (i-iv), the preprocessing steps included Nonparametric Nonuniform intensity Normalization (N4) bias field correction [47], resampling to a uniform voxel spacing [48], and z-score normalization [49] to standardize intensity distributions. Above-mentioned RF was extracted using PyRadiomics library. To ensure robust model evaluation, we adopted both 5-fold cross-validation and external validation strategies. Specifically, we reserved data from DUKE and AMBL as an external test that preserved a ~20% external test split while improving class balance. Ground-truth molecular subtypes, including luminal A, luminal B, HER2-enriched, and NBC, were obtained from associated pathology reports.

**Table 1.** Demographics of patients in different centers used in this study.

| | Item | Multicenter Datasets | | | | | |
|---|---|---|---|---|---|---|---|
| | | AMBL | DUKE | ISPY1 | ISPY2 | NACT | TCGA |
| Age | Age < 40 | 1 (3.3%) | 59 (20.6%) | 33 (19.9%) | 208 (21.2%) | 11 (23.9%) | 3 (7.5%) |
| | Age 40–59 | 19 (63.3%) | 176 (61.3%) | 115 (69.3%) | 617 (63%) | 29 (63%) | 20 (50%) |
| | Age ≥ 60 | 10 (33.3%) | 52 (18.1%) | 18 (10.8%) | 155 (15.8%) | 6 (13%) | 17 (42.5%) |
| | NA | 0 (0%) | 0 (0%) | 0 (0%) | 0 (0%) | 0 (0%) | 0 (0%) |
| Grede | High | 9 (30%) | 102 (35.5%) | 0 (0%) | 0 (0%) | 0 (0%) | 0 (0%) |
| | Intermediate | 14 (46.7%) | 80 (27.9%) | 0 (0%) | 0 (0%) | 0 (0%) | 0 (0%) |
| | Low | 5 (16.7%) | 9 (3.1%) | 0 (0%) | 0 (0%) | 0 (0%) | 0 (0%) |
| | NA | 2 (6.7%) | 96 (33.4%) | 166 (100%) | 980 (100%) | 46 (100%) | 40 (100%) |
| Molecular Subtype | Luminal A | 20 (66.6%) | 123 (42.8%) | 67 (40.3%) | 381 (38.8%) | 21 (45.6%) | 3 (7.5%) |
| | Luminal B | 3 (10%) | 49 (17%) | 25 (15%) | 155 (15.8%) | 8 (17.3%) | 32 (80%) |
| | HER2 Enriched | 2 (6.6%) | 30 (10.4%) | 29 (17.4%) | 86 (8.7%) | 6 (13%) | 5 (12.5%) |
| | Triple Negative | 5 (16.6%) | 85 (29.6%) | 45 (27.1%) | 358 (36.5%) | 2 (4.3%) | 0 (0%) |
| | NA | 0 (0%) | 0 (0%) | 0 (0%) | 0 (0%) | 0 (0%) | 0 (0%) |
| Breast Manufacturer | Confirma | 0 (0%) | 0 (0%) | 0 (0%) | 0 (0%) | 0 (0%) | 12 (30%) |
| | General Electric | 30 (100%) | 173 (60.3%) | 111 (66.9%) | 611 (62.3%) | 46 (100%) | 28 (70%) |
| | Philips | 0 (0%) | 0 (0%) | 12 (7.2%) | 117 (11.9%) | 0 (0%) | 0 (0%) |
| | SIEMENS | 0 (0%) | 114 (39.7%) | 43 (25.9%) | 252 (25.7%) | 0 (0%) | 0 (0%) |
| | NA | 0 (0%) | 0 (0%) | 0 (0%) | 0 (0%) | 0 (0%) | 0 (0%) |
| | No | 29 (96.7%) | 287 (100%) | 165 (99.4%) | 951 (97%) | 46 (100%) | 0 (0%) |
| | Yes | 1 (3.3%) | 0 (0%) | 1 (0.6%) | 29 (3%) | 0 (0%) | 0 (0%) |
| | NA | 0 (0%) | 0 (0%) | 0 (0%) | 0 (0%) | 0 (0%) | 40 (100%) |
| Ethnicity | Not hispanic or Latino | 0 (0%) | 0 (0%) | 0 (0%) | 0 (0%) | 0 (0%) | 21 (52.5%) |
| | African American | 0 (0%) | 90 (31.4%) | 28 (16.9%) | 116 (11.8%) | 2 (4.3%) | 0 (0%) |
| | American Indian | 0 (0%) | 1 (0.3%) | 0 (0%) | 0 (0%) | 0 (0%) | 0 (0%) |
| | American Indian/Alaskan Native | 0 (0%) | 0 (0%) | 0 (0%) | 4 (0.4%) | 0 (0%) | 0 (0%) |
| | Asian | 0 (0%) | 7 (2.4%) | 7 (4.2%) | 68 (6.9%) | 3 (6.5%) | 0 (0%) |
| | Caucasian | 0 (0%) | 175 (61%) | 127 (76.5%) | 777 (79.3%) | 29 (63%) | 0 (0%) |
| | Hawaian | 0 (0%) | 1 (0.3%) | 0 (0%) | 0 (0%) | 0 (0%) | 0 (0%) |
| | Hawaiian/Pacific Islander | 0 (0%) | 0 (0%) | 1 (0.6%) | 5 (0.5%) | 0 (0%) | 0 (0%) |
| | Hispanic | 0 (0%) | 6 (2.1%) | 0 (0%) | 0 (0%) | 3 (6.5%) | 0 (0%) |
| | Multiple Race | 0 (0%) | 3 (1%) | 1 (0.6%) | 7 (0.7%) | 0 (0%) | 0 (0%) |
| | Native American | 0 (0%) | 2 (0.7%) | 0 (0%) | 0 (0%) | 0 (0%) | 0 (0%) |
| | NA | 30 (100%) | 2 (0.7%) | 2 (1.2%) | 3 (0.3%) | 9 (19.6%) | 19 (47.5%) |
| Menopause Status | Post Menopause | 0 (0%) | 122 (42.5%) | 0 (0%) | 296 (30.2%) | 0 (0%) | 27 (67.5%) |
| | Pre Menopause | 0 (0%) | 164 (57.1%) | 0 (0%) | 512 (52.2%) | 0 (0%) | 13 (32.5%) |
| | NA | 30 (100%) | 1 (0.3%) | 166 (100%) | 172 (17.6%) | 46 (100%) | 0 (0%) |
| **Total** | | **30 (100%)** | **287 (100%)** | **166 (100%)** | **980 (100%)** | **46 (100%)** | **40 (100%)** |

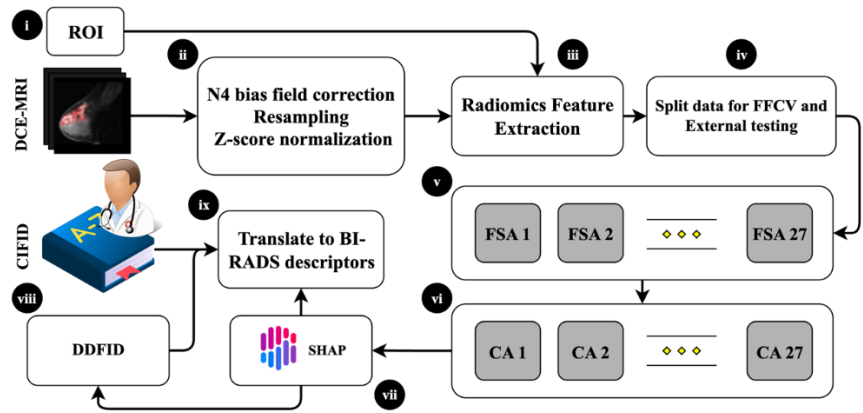

**Fig. 1.** Overview of the proposed workflow: (i) obtaining DCE-MRI scans and corresponding segmentation data from multiple centers; (ii) performing preprocessing steps including Nonparametric Nonuniform intensity Normalization (N4) bias field correction, resampling, and Z-score normalization; (iii) extracting radiomic features from the segmented regions on the preprocessed DCE-MRI using PyRadiomics; (iv) splitting the dataset for five-fold cross-validation (FFCV) and external validation; (v) applying 27 different feature selection algorithms (FSAs) on the training sets to identify the subset of top 10 feature; (vi) training 27 interpretable and complex classification algorithms (CAs) using the selected features; (vii) using SHAP analysis to assess each feature's contribution to predicting the TNBC class; (viii) constructing the Data-Driven Feature Interpretation Dictionary (DDFID) based on SHAP outputs; and (ix) translating radiomic features into BI-RADS descriptors using both the Clinically-Informed Feature Interpretation Dictionary (CIFID) and the DDFID.

***Interpretable Machine Learning (ML) Algorithms.*** For the understandable classification task, as shown in Figure 1, (v-vi), we used 27 classical and advanced classifiers, including Decision Tree Classification (DTC) [50], Logistic Regression (LOC) [51], Linear Discriminant Analysis (LDA) [52], Naive Bayes Classifier (NBC), K-Nearest Neighbors (KNN), RuleFit Classifier (RUC) [53], along with seven black-box ensemble techniques like Random Forest (RFC) [54], XGBoost Classifier (XGB), LightGBM Classifier (LGB), CatBoost Classifier (CBC [55]), Support Vector Machine (SVM) [56], Stacking Classifier (STC), and Multilayer Perceptron (MLP) [57] combined with 27 Feature Selection Algorithms (FSA) including Chi-Square Test (CST), Correlation Coefficient (CCF), Mutual Information (MIS), Variance Threshold (VTS), ANOVA F-test (AFT), Information Gain (IGS), Univariate Feature Selection (UFS), Fisher Score (FSF) [58, 59], and Least Absolute Shrinkage and Selection Operator (LAS) [60, 61], which selected 10 relevant features, to predict the molecular subtype using RF. Based on classification performance, the most important RFs were selected and translated to BI-RADS descriptors using a dictionary. The following performance metrics-Accuracy, Precision, Recall, F1-score, Receiver Operating Characteristic - Area Under the Curve (ROC-AUC), and Specificity [62, 63]-were reported as both average values and standard deviations across the five cross-validation folds and external test evaluations. Model selection was based on the highest performance across all metrics during five-fold cross-validation and externally validated using independent test sets.

***Construction of Data-Driven Feature Interpretation Dictionary (DDFID).*** To construct the DDFID, we first established a foundational link between established BI-RADS descriptors and molecular subtypes by synthesizing findings from a comprehensive literature review [64, 65, 66, 67, 68, 69, 70, 71, 72]. This review revealed a consensus profile for TNBC, which is frequently associated with descriptors such as Rim Enhancement, Round/Oval Shape, and Smooth Margins. Conversely, non-TNBC subtypes are more commonly characterized by Heterogeneous Enhancement, Irregular Shape, and Spiculated Margins. With this established framework, as shown in Figure 1 (vii-ix), we then used the SHAP analysis from our classification model to interpret the RFs that were not already defined in our CIFID. The logic was direct: if a feature's SHAP values indicated a strong positive contribution toward predicting the TNBC class, we mapped its meaning to the established TNBC descriptor profile. Similarly, if a feature contributed strongly to the non-TNBC class, it was mapped to the non-TNBC descriptor profile. Table 2 shows the relationship between molecular subtypes and BI-RADS descriptors in the context of classifying TNBC. However, for descriptors such as NME and kinetic curves, the studies we reviewed presented inconsistent or inconclusive associations with specific BI-RADS scores. Due to the limited clarity and reproducibility in these areas, we excluded NME and kinetic-related descriptors from our study.

**Table 2.** Relation between molecular subtypes and BI-RADS descriptors in the context of classifying triple-negative breast cancer (TNBC), IE: Internal Enhancement

| Type | Non-TNBC | TNBC |
|---|---|---|
| **Shape** | Irregular | Round/Oval |
| **Margin** | Spiculated | Smooth |
| **IE** | Heterogeneous | Rim Enhancement |

## 3. RESULTS

### 3.1. Relationship Between BI-RADS Scoring System and Descriptors

As shown in Table 3, The lesion's shape is a pivotal BI-RADS descriptor on breast MRI with significant prognostic implications. In the multivariate analysis by Fujiwara et al. [73], irregular shape was independently associated with higher BI-RADS categories, particularly 4 and 5, which carry increased malignancy risk and poorer outcomes. Conversely, round or oval morphology was more commonly observed in BI-RADS 2 and 3 lesions, typically reflecting benign or probably benign findings with favorable prognoses, as elaborated in the lesion-based review of BI-RADS 3 by Nguyen et al. [74]. Further, multiparametric MRI studies combining DCE-MRI, DWI, and synthetic MRI have underscored the importance of margin and IE characteristics in stratifying BI-RADS 4 lesions [75]. Lesion margins are critical BI-RADS descriptors on breast MRI that strongly correlate with malignancy risk and patient prognosis. In the Kaiser score framework, non-circumscribed margins including irregular and spiculated patterns substantially increase malignancy probability, with spiculations being highly suggestive of cancer [76]. A multivariate analysis in the BI-RADS 5th edition grading system further confirmed that irregular or spiculated margins are independently associated with BI-RADS categories 4 and 5, which carry a higher risk of malignancy and poorer outcomes [73].

Conversely, circumscribed margins are predominantly seen in BI-RADS 2 and 3 lesions, typically indicating benign or probably benign findings with favorable prognoses, as highlighted in both BI-RADS 3 reviews and Kaiser score pictorial guides. IE patterns on breast MRI serve as crucial BI-RADS descriptors that reflect lesion biology and prognostic risk. In the BI-RADS 5th edition multivariate grading analysis, heterogeneous IE, which includes non-uniform enhancement and clumping, emerged as a strong predictor of malignancy and was significantly associated with BI-RADS categories 4 and 5, indicating higher risk and poorer prognosis [77, 78]. Conversely, homogeneous enhancement, characterized by uniform, confluent contrast uptake throughout the lesion, is predominantly observed in BI-RADS 2 and 3 lesions.

**Table 3.** Studies that investigated the relationships between BI-RADS descriptors and BI-RADS scoring, IE: Internal Enhancement

| Type | Descriptor | Sematic Meaning | BI-RADS 2,3 | BI-RADS 4,5 |
|---|---|---|---|---|
| **Shape** | Irregular | Overall form is not uniform. | - | [74, 75, 79, 73] |
| | Round/Oval | Smooth, curved, and symmetrical shape. | [73, 74, 80] | [74] |
| **Margin** | Spiculated | Pointed projections from the surface. | - | [81, 76, 73, 74, 75, 79] |
| | Circumscribed | Clearly defined and sharp borders. | [82, 83, 74, 80, 73] | [76, 74] |
| | Irregular | Edge is uneven and jagged. | - | [81, 76, 74, 75, 79, 73] |
| **IE** | Heterogenous | Mixed composition of different tissues. | - | [81, 82, 73] |
| | Homogenous | Uniform composition of similar tissue. | [74, 80, 73] | - |
| | Rim Enhancement | Outer border brightens with contrast. | - | [82, 73] |
| | Dark Septations | Internal walls dividing the mass. | - | [82] |
| **Kinetics** | Plateau | Contrast enhancement level is constant. | [74, 73] | [82, 75, 79] |
| | Persistent | Contrast enhancement slowly increases. | [76, 83, 73, 74, 80, 83, 81] | [76, 75, 79] |
| | Washout | Contrast enhancement rapidly decreases. | - | [74, 73] |

### 3.2. Clinically-Informed Feature Interpretation Dictionary (CIFID)

Based on the datasets and RFs described in the previous section, we constructed a comprehensive dictionary that bridges BI-RADS imaging descriptors and standardized RFs. This dictionary integrates two key dimensions: (1) the clinical relationship between descriptors (e.g., shape, margin, and IE) and BI-RADS categories derived from literature and clinical guidelines, and (2) the visual and semantic interpretation of higher RF values, informed by the PyRadiomics documentation and relevant peer-reviewed studies [84]. To ensure clinical relevance and interpretability, this mapping was developed under the supervision and review of expert breast radiologists and oncologists. In total, we identified and linked 56 RFs that correspond meaningfully to three primary descriptor categories: shape, margin, and IE. The resulting dictionary, presented in Table 4, serves as a foundational resource for enhancing interpretability in RF-based models and aligning quantitative imaging biomarkers with familiar clinical semantics. The complete version of the CIFID, is provided in Supplemental File, Sheet 2.

**Table 4.** Clinically-Informed Feature Interpretation Dictionary (CIFID) based on expert opinion and literature review, IE: Internal Enhancement, FO: First-Order, SF: Shape-based features, GLCM: Gray Level Co-occurrence Matrix, GLSZM: Gray Level Size Zone Matrix, GLRLM: Gray Level Run Length Matrix, NGTDM: Neighborhood Gray Tone Difference Matrix, GLDM: Gray Level Dependence Matrix

| Descriptor | | Radiomic Feature | Meaning |
|---|---|---|---|
| **Shape** | Round/Oval (BI-RADS 2–3) | Flatness (3D) (F_3D) from SF | More sphere-like shape |
| | | Sphericity (3D) (Sp_3D) | Closer to perfect sphere |
| | Irregular (BI-RADS 4–5) | Perimeter to Surface ratio (PSR_2D) from SF | Greater 2D shape irregularity |
| | | Surface Area to Volume ratio (3D) (SAVR_3D) from SF | More complex or irregular surface |
| **IE** | Homogeneity (BI-RADS 2–3) | Kurtosis (Ku) from FO | Intensities cluster tightly around the mean |
| | | Uniformity (Un) from FO | Same intensity appears throughout the lesion |
| | | Autocorrelation (AC) from GLCM | Neighboring areas have similar intensity patterns |
| | | Correlation (Corr) from GLCM | Intensities follow similar trends across tissue |
| | | Cluster Tendency (CT) from GLCM | Similar pixels group together in regions |
| | | Inverse Difference (ID) from GLCM | Nearby intensities are very close together |
| | | Inverse Difference Moment (IDM) from GLCM | Smooth, gradual intensity transitions across tissue |
| | | Inverse Difference Normalized (IDN) from GLCM | Less variation in local tissue intensity |
| | | Informational Measure of Correlation (IMC2) from GLCM | Stronger dependency between pixel intensities |
| | | Joint Energy (JE) from GLCM | A few patterns dominate throughout the lesion |
| | | Maximum Probability (MP) from GLCM | One intensity pattern dominates tissue texture |
| | | Large Dependence Emphasis (LDE) from GLDM | Broad areas share same gray intensity |
| | | Long Run Low Gray Level Emphasis (LRLGLE) from GLRLM | Extended dark tissue patterns are common |
| | | Large Area Emphasis (LAE) from GLSZM | Larger regions with consistent intensity dominate |
| | | Low Gray Level Zone Emphasis (LGLZE) from GLSZM | Bigger dark zones appear in tissue |
| | | Inverse Difference Moment Normalized (IDMN) from GLCM | Texture becomes smoother and more uniform |
| | | Inverse Variance (IV) from GLCM | Pixel differences are small and stable |
| | | Coarseness (Coar) from NGTDM | Texture is broad and changes slowly |
| | Heterogeneity (BI-RADS 4–5) | Difference Average (DA) from GLCM | Greater differences between nearby tissue intensities |
| | | Informational Measure of Correlation (IMC1) from GLCM | Texture patterns are more complex overall |
| | | Maximal Correlation Coefficient (MCC) from GLCM | Structural relationships become more complex |
| | | Run Entropy (REn) from GLRLM | Intensity patterns vary more across tissue |
| | | Zone Entropy (ZE) from GLSZM | Uniform regions vary more in size |
| | | Entropy (En) from FO | Tissue intensity is more disordered overall |
| | | Interquartile Range (IQR) from FO | Intensity varies more within central range |
| | | Mean Absolute Deviation (MAD) from FO | Tissue intensities differ more from average |
| | | Range (R) from FO | Wider span of intensities in tissue |
| | | Robust Mean Absolute Deviation (rMAD) from FO | Intensity shifts even after outlier removal |
| | | Skewness (Sk) from FO | Asymmetric intensity distribution in tissue |
| | | Variance (V) from FO | Intensity values are more spread out |
| | | Contrast (Co) from GLCM | Strong local differences in tissue brightness |

| | |
|---|---|
| Cluster Prominence (CP) from GLCM | Sharp intensity differences in clustered regions |
| Cluster Shade (CS) from GLCM | Unevenness in tissue intensity clustering |
| Difference Entropy (DiEn) from GLCM | Intensity differences are more irregular |
| Joint Entropy (JEn) from GLCM | Gray-level combinations are more diverse |
| Sum Entropy (SEn) from GLCM | Tissue brightness combinations are more irregular |
| Sum of Squares (SQ) from GLCM | Tissue values deviate further from mean |
| Gray Level Non-Uniformity (GLN) from GLDM | Some intensities occur more than others |
| Small Dependence Emphasis (SDE) from GLDM | Many small uniform regions in tissue |
| Small Dependence High Gray Level Emphasis (SDHGLE) from GLDM | Bright small regions appear more often |
| Small Dependence Low Gray Level Emphasis (SDLGLE) from GLDM | Dark small regions appear more often |
| Gray Level Non-Uniformity (GLN) from GLSZM | Brightness varies across tissue zones |
| Gray Level Non-Uniformity Normalized (GLNN) from GLSZM | More brightness variation between regions |
| Gray Level Variance (GLV) from GLSZM | Intensities differ more across zones |
| Small Area Emphasis (SAE) from GLSZM | Many small, similar-intensity tissue patches |
| Size-Zone Non-Uniformity (SZN) from GLSZM | Zones differ more in tissue structure |
| Size-Zone Non-Uniformity Normalized (SZNN) from GLSZM | Normalized zone size variation increases |
| Zone Percentage (ZP) from GLSZM | Numerous small texture zones in lesion |
| Zone Variance (ZV) from GLSZM | Greater spread in tissue zone sizes |
| Complexity (Com) from NGTDM | Tissue intensity varies in complex patterns |
| Busyness (B) from NGTDM | Rapid changes in local tissue intensity |
| Contrast (Co) from NGTDM | Sharp brightness transitions between neighboring areas |

### 3.3. Understandable Classification Task

***Subtype classification performance.*** Among the evaluated models, those utilizing the Extra Trees classifier demonstrated the highest performance in distinguishing molecular subtypes. The best-performing model, which employed the VIF feature selection method, achieved a validation accuracy of $0.83 \pm 0.02$ and an external testing accuracy of $0.67 \pm 0.01$. This was closely followed by the model using UFS, which achieved a validation accuracy of $0.82 \pm 0.02$ and a test accuracy of $0.69 \pm 0.008$, and the model using MIGR, which also reached a validation accuracy of $0.82 \pm 0.02$ and a test accuracy of $0.67 \pm 0.01$. Figure 2 presents a heatmap of validation accuracies across five representative classifiers. These include Extra Trees, Bagging, and XGBoost as high-performing models, GNB as the lowest-performing classifier, and RF as a model with intermediate performance. Comprehensive results—including average ROC-AUC, Precision, Recall, F1-score, Specificity, and Accuracy—for five-fold cross-validation and external testing across all remaining classifier and feature selection combinations are provided in Supplemental File, Sheet 3. Additionally, Sheets 4 and 5 detail the classifier hyperparameters and the top features selected by the respective FSAs.

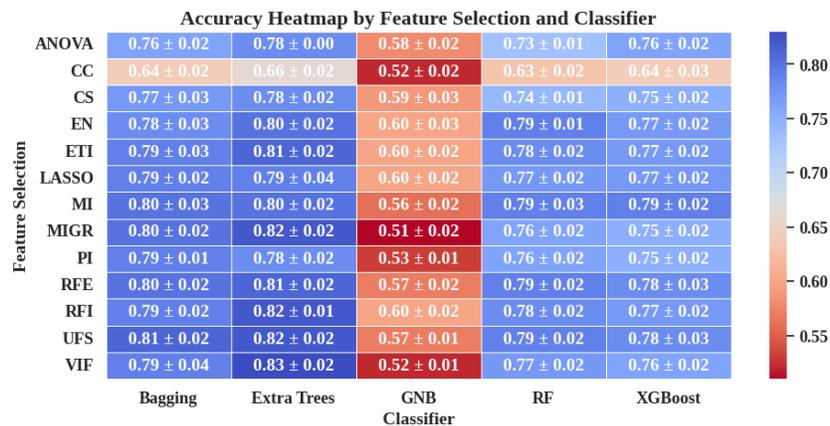

**Fig 2.** Performance summary of 5 classifiers combined 13 feature selection algorithms, EN: ElasticNet, RFE: Recursive Feature Elimination, ETI: Extra Trees Importance, LASSO: Least Absolute Shrinkage and Selection Operator, MIGR: Mutual Information Gain Ratio RFI: Feature Importance RF, CS: Chi Square Test, ANOVA: Analysis of Variance Test, MI: Mutual Information, UFS: Univariate Feature Selection PI: Permutation Importance, VIF: Variance Inflation Factor, CC: Correlation Coefficient, GNB: Gaussian Naive Bayes, RF: Random Forest

***Data-Driven Feature Interpretation Dictionary (DDFID).*** Using SHAP analysis on the trained classification model, we identified the RFs most influential in distinguishing TNBC from non-TNBC cases as follows in the Supplemental File, Sheet 6. By aligning the direction and magnitude of SHAP values with the clinical BI-RADS descriptors established in our literature review, we derived a semantic interpretation for previously undefined 52 RFs and incorporated them into our DDFID. Features such as Sum Average from GLCM and Voxel Volume from SF that showed consistently high SHAP values contributing toward the TNBC class were semantically mapped to descriptors such as Rim Enhancement, Round/Oval Shape, and Smooth Margins as characteristics previously associated with TNBC subtypes. Conversely, features that contributed strongly to non-TNBC predictions, such as 10 Percentile form FO and Large Area Low Gray Level Emphasis from GLSZM were linked to descriptors like Heterogeneous Enhancement, Irregular Shape, and Spiculated Margins, which dominate the imaging profile of other molecular subtypes. This mapping allowed us to extend the existing CIFID with a data-driven layer of interpretability, grounded in model behavior. The complete version of the DDFID, is provided in Supplemental File, Sheet 7.

*Feature interpretation.* We used the above-mentioned SHAP values to assess the contribution of each RF in predicting the TNBC class. As shown in Figure 3, blue-colored SHAP values indicate that higher values of a given feature are associated with a higher likelihood of belonging to the TNBC group, while red values suggest that higher values of that feature are associated with a lower probability of being TNBC. To interpret these abstract features in a clinically meaningful way, we used both our CIFID and DDFID to map the RF terms to their corresponding morphological descriptors. For instance, the feature Sphericity from the shape category (SF_Sp) typically reflects the roundness of a tumor, with higher values indicating a more spherical or oval morphology and smoother margins.

In our model, elevated SF_Sp values were strongly associated with the TNBC subtype, suggesting that TNBC tumors in our dataset tend to exhibit more rounded shapes with circumscribed margins. Another important feature, busyness (B), quantifies the local intensity variation within the tumor. Higher values generally indicate greater heterogeneity. However, in our results, lower busyness values were predictive of TNBC, implying a more homogeneous IE pattern. This observation aligns with findings from previous studies, which have reported more uniform enhancement in TNBC lesions on dynamic contrast-enhanced MRI. Interestingly, although the feature Flatness (F) is often interpreted such that higher values suggest a more ellipsoidal or flat morphology, our model associated lower Flatness values, indicating more irregular shapes with TNBC. This appears counterintuitive and may suggest either a misrepresentation by the model or an underlying morphological pattern that is not yet fully understood. While such findings warrant further validation, they demonstrate how feature interpretation must be grounded in both RF theory and clinical context to derive biologically meaningful insights. Supplemental File, Sheets 8 and 9, indicates feature importance values for both CIFID and DDFID.

**CIFID**

| Feature | FS | Effect On TNBC | Meaning |
|---|---|---|---|
| Flatness (SF_F_3D) | FIRF | (red, negative) | S: Irregular |
| Skewness (FO_Sk) | FIRF | (red, negative) | IE: Homogenous |
| Large Area Emphasis (GLSZM_LAE) | UFS | (red, negative) | IE: Heterogenous |
| Sphericity (SF_Sp_2D) | FIRF | (blue, positive) | S: Round/Oval |
| Small Dependence Emphasis (GLDM_SDE) | FIRF | (blue, positive) | IE: Heterogenous |
| Large Area Emphasis (GLSZM_LAE) | MIGR | (red, negative) | IE: Heterogenous |
| Busyness (NGTDM_B) | MIGR | (red, negative) | IE: Homogenous |
| Zone Variance (GLSZM_ZV) | MIGR | (blue, positive) | IE: Heterogenous |
| Gray Level Non-Uniformity (GLSZM_GLN) | MIGR | (blue, positive) | IE: Heterogenous |
| Large Dependence Emphasis (GLDM_LDE) | UFS | (red, negative) | IE: Heterogenous |

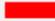

**Fig. 3.** Important features identified in our interpretable classification task resulted from Extra Tree Classifier as our best classifier, along with their corresponding translations: features mapped using the Clinically-Informed Feature Interpretation Dictionary (CIFID), and those interpreted using the Data-Driven Feature Interpretation Dictionary (DDFID). FS: Feature Selection, FIRF: Feature Importance Random Forest, MIGR: Mutual Information Gain Ratio, UFS: Univariate Feature Selection, IE: Internal Enhancement, S: Shape, M: Margin, TNBC: Triple Negative Breast Cancer

## 4. DISCUSSION

We developed a new method to interpret RFs from breast MRI using two complementary dictionaries. The first, CIFID, links 56 features to BI-RADS descriptors based on literature, physicians' expertise, and PyRadiomics definitions. The second, DDFID, used statistical analysis from a prediction model (TNBC vs. non-TNBC) to interpret the remaining features. Our approach to developing foundational dictionaries offers a distinct advantage over common interpretability methods in RFs. This methodology is part of a broader effort by our team to create standardized, translatable frameworks, as demonstrated by our parallel work in prostate MRI, where we developed a similar dictionary (PM1.0) to connect PI-RADS semantic features with quantitative data [85], and in liver cancer pathology, where we introduced the Pathobiological Dictionary (LCP1.0) to map WHO grading criteria to RFs and pathomics features for interpretable AI-driven diagnostics [86]. In contrast, other prevalent strategies yield findings that are often confined to a single experiment. Post-hoc analysis, for instance, Duan et al [87] explain a model after its creation using techniques like SHAP, as demonstrated by Luo et al [88] in their work predicting lymph node metastasis in non-small cell lung cancer.

Other powerful approaches link specific RFs to underlying biological truth, such as correlating MRI signatures with histological grade as demonstrated by Wang et al [89], or linking imaging phenotypes to tumor gene expression as shown by Bismeijer et al [90]. While these methods provide valuable, model-specific explanations by connecting imaging to pathology and radiogenomics, they do not typically result in a comprehensive, reusable dictionary. Our focus, therefore, is on creating such a generalizable resource. The developed CIFID is designed as a reusable, standardized reference that can be validated and improved across future studies, providing a more scalable and clinically translatable path for understanding RF models.

Our study's performance, achieving a test accuracy of 0.69 in distinguishing triple-negative from non-triple-negative breast cancer, is competitive within the field, particularly considering our rigorous methodology. For comparison, a study by Demircioglu et al. [91] reported a similar though slightly higher performance for the same classification task, achieving an ROC-AUC of 0.73. The approach, however, was validated on a much smaller, single-center dataset of 98 patients. Another relevant study by Ma et al. [92], developed a RF model that reported a higher ROC-AUC of 0.87 on their independent test set, although this was derived from a cohort of only 81 patients. While our test accuracy is positioned between these findings, our model's development on a significantly larger and more diverse dataset, combined with our comprehensive multi-center validation, suggests a greater degree of generalizability.

An analysis of the most predictive features from our top three classifiers reveals a complex but insightful relationship with the established literature. While our literature-derived CIFID accounted for 7 of the 30 most important features, the majority were identified through our data-driven DDFID approach. Comparing our model's findings to the literature shows areas of both consistency and divergence. For instance, our models identified shape features as highly predictive for TNBC, which aligns with reports that TNBC often presents as round or oval masses with circumscribed margins [64, 65, 68, 70]. However, the specific contribution of any single feature in our model does not always perfectly mirror its reported importance in previous studies. For example, while several of the above-mentioned studies highlight 'round/oval' as a key descriptor for TNBC, our model prioritized texture features like shape_Flatness as a positive contributor to TNBC or sometimes shape_sphericity as a negative contributor to TNBC. Although many previous studies [65, 70] defined "heterogenous" IE as characteristic of non-TNBC, our model detected Gray Level Non-Uniformity (GLN) from GLDM (GLDM_GLN) as a positive contributor to TNBC. These discrepancies are expected and can be attributed to three key factors. First, our model's performance, while strong, is not absolute. A machine learning model's feature selection is driven by optimizing predictive accuracy on a specific dataset.

A biologically relevant feature may not be selected if it does not sufficiently improve the model's performance, or its importance might be overshadowed by other correlated features. Second, the existing literature on imaging descriptors for molecular subtypes is not definitive and often lacks a clear consensus. Third, for the features that are translated using our developed DDFID, these discrepancies might stem from their uncertain statistical nature. Different studies report varied, and sometimes conflicting, associations, as highlighted in systematic reviews [39], meaning there is no perfect "gold standard" for comparison. This complex landscape underscores the value of our approach. By identifying features that are important for prediction, we can highlight potential biomarkers that may not be emphasized in the ambiguous literature. The goal is not a perfect one-to-one match, but to use a data-driven methodology to bring clarity and identify the most robust predictors, paving the way for a more refined understanding of the imaging phenotypes of breast cancer subtypes.

There are several limitations to our work. The CIFID was developed using a limited number of physicians and lacks validation across institutions or populations. The statistical dictionary is based on a model with imperfect generalization performance, meaning the feature-subtype associations it produces may not be entirely reliable. Additionally, although we initially attempted to build a four-class classifier to distinguish between all molecular subtypes, the imbalanced data and small sample sizes in some groups led to unsatisfactory performance, so we focused on binary classification instead. Finally, the statistical dictionary is inherently dependent on the performance of the classifier, meaning its quality improves as the model becomes more accurate.

This study has several limitations that warrant future exploration. Broader validation with more clinicians is needed to refine the clinical dictionary, while larger, more diverse datasets will improve generalizability. Exploring alternative model architectures and feature selection strategies may enhance predictive performance [93, 94]. Integrating RFs with other modalities, such as transcriptomics or digital pathology, could yield more robust and biologically meaningful models. Additionally, prior studies highlight the importance of assessing RF stability across different segmentations and imaging conditions, as variability can impact reliability. Future work also aims to expand the framework using ~ 500 RFs extracted through PySERA [95], enabling more standardized, explainable, and reproducible pipelines. Prospective validation and real-time clinical feedback will be key for successful clinical integration.

## 5. CONCLUSION

In this study, we introduced a novel dual-dictionary approach (CIFID and DDFID) to make RFs more interpretable for classifying triple-negative breast cancer. Our model achieved acceptable predictive performance by confirming known semantic features, like tumor shape, while also uncovering novel, quantitative texture features as powerful biomarkers. The primary contribution is to introduce a reusable and extensible framework that addresses the "black box" challenge in AI-guided clinical assessments, moving beyond study-specific interpretations. Our methodology highlights the need to augment the standard radiological lexicon with data-driven insights to capture the subtle heterogeneity crucial for accurate classification. While this retrospective study has limitations, future work will focus on prospective validation and expanding the dictionaries to other cancer subtypes. This work represents a significant step toward developing more trustworthy and clinically meaningful AI tools to advance personalized medicine in breast cancer care.

**Data, Machine Learning Hyperparameters, and Code Availability.** All codes and supplemental files are publicly shared at: https://github.com/MohammadRSalmanpour/Radiological-and-Biological--Radiomics-Feature--based-Dictionary-Version-BM1.0

**Acknowledgment.** This study was supported by the Virtual Collaboration Group (VirCollab, www.vircollab.com), and the Technological Virtual Collaboration (TECVICO CORP.) based in Vancouver, Canada. We gratefully acknowledge funding from the Canadian Foundation for Innovation – John R. Evans Leaders Fund (CFI-JELF; Award No. AWD-023869 CFI), as well as the Natural Sciences and Engineering Research Council of Canada (NSERC) Awards AWD-024385, RGPIN-2023-0357, and Discovery Horizons Grant DH-2025-00119.

**Conflict of Interest.** The co-author, Mohammad R. Salmanpour, is affiliated with TECVICO CORP. The remaining co-authors declare no relevant conflicts of interest.